\documentclass[prb,aps,twocolumn,showpacs,floatfix]{revtex4-2}

\usepackage{graphicx,color}
\usepackage{amsthm}
\usepackage{amsfonts}
\usepackage{algorithmic}
\usepackage{enumerate}
\usepackage{latexsym}
\usepackage{amsmath}
\usepackage{amssymb}
\usepackage{bm}
\usepackage[pdftex,plainpages=false,colorlinks=true,linkcolor=blue, citecolor=blue, urlcolor=blue]{hyperref}
\emergencystretch=\maxdimen
\begin{document}
\title{The odd-parity altermagnetism induced reconstruction of the Chern-insulating phase in Haldane-Hubbard model}

\author{Minghuan Zeng$^{1}$}
\author{Zheng Qin$^{1}$}
\author{Ling Qin$^{2}$}
\author{Shiping Feng$^{3,4}$}
\author{Jia-Cheng He$^{5}$}
\email{jche14@fudan.edu.cn}
\author{Dong-Hui Xu$^{1,6}$}
\email{donghuixu@cqu.edu.cn}
\author{Rui Wang$^{1,6}$}
\email{rcwang@cqu.edu.cn}

\affiliation{$^{1}$Institute for Structure and Function \& Department of Physics \& Chongqing Key Laboratory for Strongly Coupled Physics,
Chongqing University, Chongqing, 400044, People's Republic of China}

\affiliation{$^{2}$College of Physics and Engineering, Chengdu Normal University, Chengdu, 611130, Sichuan, People's Republic of China}

\affiliation{$^{3}$Department of Physics, Faculty of Arts and Science, Beijing Normal University, Zhuhai, 519087, People's Republic of China}

\affiliation{$^{4}$School of Physics and Astronomy, Beijing Normal University, Beijing, 100875, People's Republic of China}

\affiliation{$^{5}$Centre for Modern Physics, Chongqing University, Chongqing 400044, People's Republic of China}

\affiliation{$^{6}$Center of Quantum materials and devices, Chongqing University, Chongqing 400044, People's Republic of China}

\begin{abstract}
Odd-parity altermagnetism(ALM) extends compensated collinear magnetism beyond the even-parity spin splitting of conventional altermagnets,
but its role in correlated topological phases remains largely unexplored. Using the cluster slave-spin method, we show that the odd-parity ALM appearing
in the ALM Chern-insulating phase of Haldane-Hubbard model significantly reconstructs the local distribution of the Berry curvature 
in the conventional Chern-insulating phase, while the total Chern number remains unchanged compared to the Chern-insulating phase.
The Berry curvature becomes spin and valley selective; zigzag ribbons develop nonsymmetric chiral edge states;
while armchair ribbons remain inversion symmetric. The optical response mirrors this separation between the local reconstruction and the global topology:
low-energy spectra are governed by quasiparticles near the gap,
whereas the low-frequency Hall conductivity stays quantized, $\sigma_{\rm T\uparrow}(\Omega\to 0)=\sigma_{\rm T\downarrow}(\Omega\to 0)=e^2/h$.
These results establish the Haldane-Hubbard model as a minimal correlated platform for odd-parity altermagnetic topology.
\end{abstract}


\maketitle


\section{Introduction}\label{Introduction}

Altermagnets (ALMs) have recently been identified as the third fundamental class of collinear magnetism, distinct from both ferromagnets and conventional antiferromagnets\cite{Libor22_1,Libor22_2,Bai24}. Their compensated real-space order coexists with momentum-dependent spin splitting, enabling time-reversal-breaking responses without net magnetization. This principle has rapidly expanded the scope of spintronics and topological transport, especially in even-parity ALMs where anomalous Hall transport, charge-to-spin conversion, and related Berry-curvature effects have already been established\cite{Libor20,Rafael21,Shao21,Karube22,Feng22,HuM25}. A major current direction is to move beyond the even-parity setting and understand odd-parity nonrelativistic spin splitting, which provides a magnetic analogue of Dresselhaus-type physics and opens a wider symmetry landscape for altermagnetic transport and topology\cite{Manchon15,Hellenes24,Song25,Yu25,Brekke24,Wang25}.

This problem is particularly important in correlated systems. The symmetry analysis on the basis of spin group theory\cite{Brinkman66,Litvin74,Litvin77}
has shown how odd-parity altermagnetism can emerge in compensated collinear magnets when opposite-spin sublattices are connected by symmetries
such as $[C_2||\bar{E}]$ or $[C_2||M]$\cite{Zeng25_1}. At the same time, recent APS studies have emphasized that altermagnetic responses
can be strongly shaped by Berry curvature, electronic correlations, and quantum geometry\cite{Sato24,Fang24}.
The key open question is therefore not only whether an odd-parity altermagnetic phase exists, but also how it reorganizes momentum-space topology,
edge spectra, and Hall transport, especially in the presence of interactions.

The Haldane-Hubbard (HH) model provides a natural minimal platform for this question. It is a paradigmatic interacting Chern-insulator model, and recent works have shown that,
at large Haldane hopping and intermediate interaction strength, it hosts an odd-parity altermagnetic Chern-insulating(ALM-CI) phase\cite{Lin25,Zeng25_1}. Previous studies for the HH model mainly focused on phase boundaries, bulk gaps, and changes of the Chern invariant\cite{Zheng15,Arun16,Wu16,Vanhala16,ifmmode16,Mertz19,He24,Mai23}, whereas edge-state phenomenology is much better understood in the noninteracting Haldane limit\cite{Haldane88,Hao08}. It is still unclear how odd-parity altermagnetic order reconstructs Berry geometry and edge states when the global Hall topology remains intact.

In this work, we study this issue using the cluster slave-spin method. It is shown that the transition from the conventional Chern-insulating(CI) phase into the odd-parity ALM-CI phase leaves the total Chern number unchanged, but strongly reshapes the local distribution of the Berry curvature: the Berry curvature exhibits significant spin-valley locking; zigzag ribbons develop nonsymmetric chiral edge states; and armchair ribbons remain inversion symmetric. Using the Kubo formula\cite{Mahan90}, we further show that the energy dependence of optical response is governed by low-energy quasiparticles near the single-particle gap, whereas the low-frequency Hall conductivity remains quantized at $\sigma_{T}=e^2/h$ for each spin. The HH model thus emerges as a minimal correlated setting in which the global Chern topology survives in spite of a pronounced reconstruction of local Berry geometry, providing a concrete bridge between odd-parity altermagnetism, correlated Hall physics, and quantum-geometric transport.

The remainder of this paper is organized as follows. In Sec. \ref{Formalism}, we introduce the model and the cluster slave-spin formalism. In Subsection \ref{Results-EDisp}, we present the electronic properties, including the Berry curvature and the edge states of zigzag and armchair ribbons. In Subsection \ref{Transport}, we discuss the optical conductivities. Finally, we summarize our main results in Sec. \ref{conclude}.

\section{Formalism}\label{Formalism}

\subsection{The cluster slave spin method}

The quantum Monte Carlo(QMC) method has proven successful for half-filled bipartite lattices where the particle-hole symmetry is preserved.
However, its predictions become questionable due to the notorious sign problem for doped fermion systems and extended Hubbard models
such as the $t$-$t'$-$U$ model\cite{Duffy97,Avella97,Avella01,Fr_sard98} and the Haldane-Hubbard model\cite{Zheng15,Arun16,Wu16,Vanhala16,ifmmode16,Mertz19,He24,Mai23,Lin25,Zeng25_1}, as well as finite-size effects.
In such challenging regimes particularly at moderate coupling strengths where conventional perturbation theory breaks down,
nonperturbative techniques based on slave-variable representations have been introduced to provide physical insights.
Among these, the slave-boson approach has been widely applied to both single-band\cite{Kotliar_1986} and multiorbital systems\cite{Lechermann07}.
As shown by Kotliar and Rukenstein\cite{Kotliar_1986}, the saddle-point approximation within this framework reproduces the Gutzwiller results
for the half-filled paramagnetic state, while fluctuations beyond the saddle point capture high-energy excitations which would be important
in the moderate and high temperature regime.
Nevertheless, as the number of degrees of freedom grows, the bosonic slave variables proliferate exponentially,
rendering the slave-boson method impractical for multiorbital systems and cluster mean-field theories.

To circumvent this difficulty, the slave-spin approach\cite{Serge_2004,Zhao07,Florens02,Medici05}
was developed for the half-filled multiorbital Hubbard model, in which a slave spin represents the charge sector and an auxiliary fermionic field
accounts for the spin sector. Since only one slave spin variable is required per degree of freedom, the previously mentioned exponential bottleneck is overcome.
Later, Hassan and de' Medici\cite{Hassan2010} extended this method to non-half-filled fermion systems by developing the $Z_2$ slave-spin theory.
However, this $Z_2$ version fails to recover the noninteracting dispersion of multiorbital Hamiltonians due to nonzero orbital-dependent Lagrange multipliers.
In contrast, the $U(1)$ slave-spin theory\cite{Yu.Rong_2012} not only reproduces the $Z_2$ results in the presence of on-site interactions,
but also correctly yields the noninteracting dispersion by introducing an extra orbital-dependent effective chemical potential,
which reduces to the Lagrange multiplier at $U=0$ (see Appendix C of Ref.~\onlinecite{Zeng21} for details).
More recently, a generalized cluster slave-spin method\cite{WCLee17,Zeng21,Zeng22} has successfully described the crossover behavior
of the antiferromagnetic gap $\Delta_{\text{AFM}}$ from the weak- to strong-interaction regime and given rise to the quasiparticle residue
in agreement with the extended Gutzwiller factor\cite{TohruOgawa}.

Along with the charge-spin separation picture for strongly correlated systems\cite{PALee06,Barnes76,Coleman84,Kotliar_1986,Serge_2004} 
in the U(1) slave-spin method\cite{Yu.Rong_2012}, an electron operator is factorized into the direct product of a slave-spin operator
($S = \tfrac{1}{2}$) and a fermionic spinon operator, respectively describing its charge and spin degree of freedom, i.e.,
\begin{equation}\label{s-spin-representation}
C^{\dagger}_{\alpha}\equiv S^{+}_{\alpha}f^{\dagger}_{\alpha}\;,
\end{equation}
on account of which the original Hilbert space of an electron with the basis $\{|0\rangle, |1\rangle\}$ is enlarged to
$\{|n^{f}_{\alpha},S^{z}_{\alpha}\rangle\} = \{|0,-\tfrac{1}{2}\rangle,|1,\tfrac{1}{2}\rangle, |0,\tfrac{1}{2}\rangle,|1,-\tfrac{1}{2}\rangle\}$.
Thus, an extra constraint needs to be enforced to restrict the Hilbert space to the physical one:
$\{|n^{f}_{\alpha},S^{z}_{\alpha}\rangle\}=\{|0,-\tfrac{1}{2}\rangle,|1,\tfrac{1}{2}\rangle\}$, i.e.,
\begin{equation}\label{constraint}
S^{z}_{\alpha}=f^{\dagger}_{\alpha}f_{\alpha}-\tfrac{1}{2}\;.
\end{equation}
A gauge degree of freedom could be introduced to impose the constraint, signifying that the electron operator in the slave-spin
representation is invariant under a local gauge transformation $f^{\dagger}_{\alpha} \to f^{\dagger}_{\alpha}
e^{-i \phi_{\alpha}}$ and $S^{+}_{\alpha} \to S^{+}_{\alpha} e^{i \phi_{\alpha}}$, and all
physical quantities should be invariant under this U(1) gauge transformation\cite{Coleman_1987,PatrickA_1992,Feng_1993,Serge_2004,Senthil_2008}.
Therefore, the charge-spin separation has been faithfully enforced, i.e., the slave spins carry the U(1) charge similarly as the slave rotors\cite{Serge_2004},
while the spinons are represented by the fermionic fields $f_{\alpha}$. 
Otherwise, following the common practice\cite{Yu.Rong_2012,Hassan2010,WCLee17,Zeng21,Zeng22}, the above local constraints can be dealt with
by introducing the local Lagrange multipliers $\lambda_{i\sigma}$.

\subsection{Model and its cluster slave-spin solution}\label{HH-Cluster-SSpin}

\begin{figure}[h!]
\includegraphics[scale=0.45]{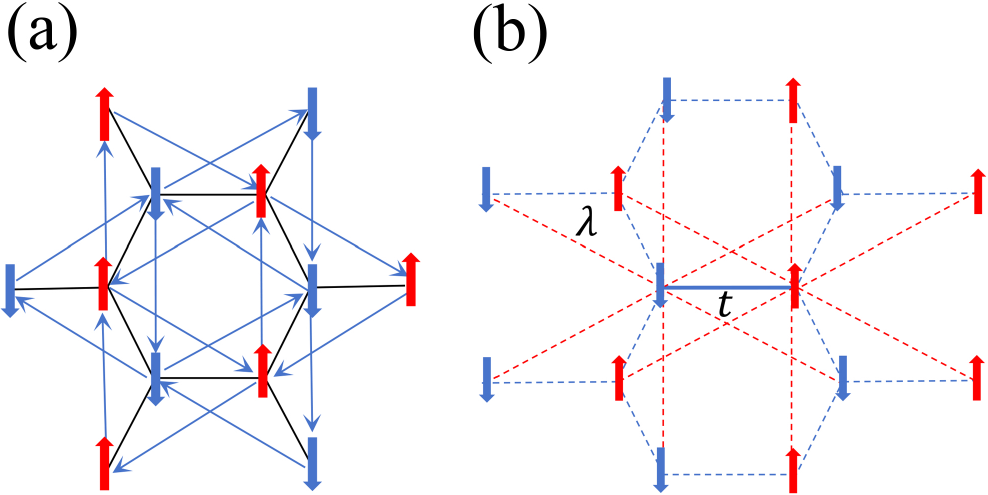}
\caption{(Color online) (a)The schematic illustration of the ground state of hexagon-lattice HH model having the compensated collinear magnetism
with opposite sublattice currents denoted by slight blue arrows, where the bold up-(red) and downward(blue) solid arrows
represent the corresponding spin polarizations.
(b)The schematic illustration of the 2-site cluster slave spin approximation, where the NN hopping denoted by the blue solid line
and the onsite electron Coulomb repulsion within the 2-site cluster are treated using the numerical method of exact diagonalization,
while the coupling between this 2-site cluster and the environment, denoted by the red and blue dashed lines, are roughly treated via the mean-field decoupling.
\label{subcurrent-2site-cluster} }
\end{figure}
In the previous work\cite{Zeng25_1}, we started from the spin group theory to derive sufficient conditions for the emergence of odd-parity ALM,
which has been further verified in square- and hexagon-lattice HH models\cite{Lin25,Zeng25_1} and in other theoretical settings\cite{Zhuang25,Luo25,Zhu26}.
In particular, the phase diagram of the hexagon-lattice HH model exhibits a topologically nontrivial odd-parity ALM-CI phase
in the intermediate interaction regime and at large Haldane hoppings, which is changed into the conventional odd-parity altermagnetic insulating(ALMI) phase
with the further increment of interaction strength. However, the relevant topological properties such as the Berry curvature
and edge states have not yet been explored in the presence of strong electronic correlations.
Therefore, here we focus on revealing the distinctive topological features of HH model in the ALM-CI phase,
in contrast to the conventional CI phase appearing at small interaction strengths. Below, we start from the HH model [cf. Fig.\ref{subcurrent-2site-cluster}(a)],
which reads
\begin{eqnarray}\label{HH-Model}
{\rm H} &=&\sum_{\langle ij\rangle}\big[ -tC_{iA}^{\dagger}C_{jB}+\textsc{h.c.} + \mu\delta_{ij} \sum_{s}C_{is}^{\dagger}C_{js} \big] \nonumber\\
 &+& \lambda\sum_{s=A,B}\sum_{\langle\langle ij\rangle\rangle}C_{is}^{\dagger}e^{i\tfrac{\pi}{2}\nu_{ij}}C_{js}
 + U\sum_{i,s=A,B}n_{is\uparrow}n_{is\downarrow}\;,
\end{eqnarray}
where $\langle ij\rangle$ and $\langle\langle ij\rangle\rangle$ denote the summation over all the nearest-(NN) and next-nearest-neighbor(NNN) sites, respectively;
$C_{is}^{\dagger}=(C_{is\uparrow}^{\dagger},C_{is\downarrow}^{\dagger})$ a two-component spinor with $s=A,B$ denoting two sublattices;
$\nu_{ij}=\pm 1$ the Haldane phase factor for clockwise and anticlockwise path connecting the NNN sites $i$ and $j$;
$n_{is\sigma}=C_{is\sigma}^{\dagger}C_{is\sigma}$ the electron occupation number operator at site $is$ with spin $\sigma$,
and $\mu$ the chemical potential introduced to determine the charge carrier doping concentration.
In the following segments, the NN hopping integral $t$ and the lattice constant $a$ are set as the energy and length unit, respectively.

Within the U(1) slave-spin representation, HH model can be recast into the form
\begin{eqnarray}\label{HH-SSPIN}
{\rm H}_{\rm S-Spin} &=&\sum_{\langle ij\rangle\sigma}\big[ -tf_{iA\sigma}^{\dagger}f_{jB\sigma}S_{iA\sigma}^{+}S_{jB\sigma}^{-} + \textsc{h.c.}\nonumber\\
&+& \delta_{ij}\sum_{s}(\mu-\lambda_{is\sigma})f_{is\sigma}^{\dagger}f_{js\sigma} \big]\nonumber\\
&+& \lambda\sum_{s=A,B}\sum_{\langle\langle ij\rangle\rangle\sigma}e^{i\tfrac{\pi}{2}\nu_{ij}}
 f_{is\sigma}^{\dagger}f_{js\sigma}S_{is\sigma}^{+}S_{js\sigma}^{-}\;,\nonumber\\
 &+&\sum_{is\sigma}\lambda_{is\sigma}(S_{is\sigma}^{z}+\frac{1}{2})\nonumber\\
 &+& U\sum_{i,s=A,B}[S_{is\uparrow}^{z}+\frac{1}{2}][S_{is\downarrow}^{z}+\frac{1}{2}]\;,
\end{eqnarray}
where the Lagrangian mulpliers $\lambda_{is\sigma}$ are introduced to faithfully implement the local constraint that restricts the enlarged Hillbert space
into the physical one,
\begin{equation}\label{local-constraint}
S^{z}_{is\sigma}=f^{\dagger}_{is\sigma}f_{is\sigma}-\tfrac{1}{2}\;.
\end{equation}
Along with previous studies\cite{Yu.Rong_2012,Hassan2010} where the mean-field decoupling between the fermionic spinon
and the slave spin has been implemented, on account of which researchers obtained the Mott insulating phase in a good agreement with the dynamical mean field theory,
and the quasiparticle weight $Z$ obtained in the large
$U$ limit reproduces the famous Gutzwiller approximation $g_t =2x/(1 +x)$\cite{Hassan2010,Gutzwiller63,Gebhard90},
the above HH Hamiltonian in the slave spin representation can be directly decomposed into the slave spin and fermionic spinon section,
\begin{subequations}\label{HH-Model-SSpin}
\begin{eqnarray}
{\rm H}_{MF}^{f} &=&\sum_{\langle ij\rangle\sigma}\big[ -tf_{iA\sigma}^{\dagger}f_{jB\sigma}\langle S_{iA\sigma}^{+}S_{jB\sigma}^{-}\rangle + \textsc{h.c.}\nonumber\\
&+& \delta_{ij}\sum_{s}(\mu-\lambda_{is\sigma})f_{is\sigma}^{\dagger}f_{js\sigma} \big]\nonumber\\
&+& \lambda\sum_{s=A,B}\sum_{\langle\langle ij\rangle\rangle\sigma}e^{i\tfrac{\pi}{2}\nu_{ij}}
 f_{is\sigma}^{\dagger}f_{js\sigma}\langle S_{is\sigma}^{+}S_{js\sigma}^{-}\rangle\;,\\
 {\rm H}_{MF}^{S} &=&\sum_{\langle ij\rangle\sigma}\big[ -t\langle f_{iA\sigma}^{\dagger}f_{jB\sigma}\rangle S_{iA\sigma}^{+}S_{jB\sigma}^{-} + \textsc{h.c.} \big]\nonumber\\
&+& \lambda\sum_{s=A,B}\sum_{\langle\langle ij\rangle\rangle\sigma}e^{i\tfrac{\pi}{2}\nu_{ij}}\langle f_{is\sigma}^{\dagger}f_{js\sigma}\rangle S_{is\sigma}^{+}S_{js\sigma}^{-}\nonumber\\
 &+&\sum_{is\sigma}\lambda_{is\sigma}(S_{is\sigma}^{z}+\frac{1}{2})\nonumber\\
 &+& U\sum_{i,s=A,B}[S_{is\uparrow}^{z}+\frac{1}{2}][S_{is\downarrow}^{z}+\frac{1}{2}]\;,
\end{eqnarray}
\end{subequations}
where $\langle \cdots \rangle$ denotes the mean-field value. We note that upon the mean-field decoupling, the dynamics of the fermionic spinons
and the bosonic slave spins are determined mainly by the spinon Hamiltonian ${\rm H}_{MF}^{f}$ and the slave spin Hamiltonian ${\rm H}_{MF}^{S}$,
respectively, while they are connected by the mean-field variables $\langle S_{is\sigma}^{+}S_{js'\sigma}^{-}\rangle$
and $\langle f_{is\sigma}^{\dagger}f_{js'\sigma}\rangle$ with the sublattice index $s, s' =$ A and B, respectively.
It is worth noting that the dynamical coupling between the spinon fields and the slave spins is
neglected within the mean-field decoupling treatment, which can be included by incorporating the vertex correction coming from the interaction between the
spinon fields and slave spins[See Eq.\eqref{HH-SSPIN}]. However, this interaction-induced vertex corrections are not supposed to cause significant impacts
on the ground-state properties of HH model, because the long-range staggered magnetization opens a gap in the spinon energy dispersion which increases the
energy cost for the coupling between the fermionic spinon fields and slave spins. 

Furthermore, the slave spin operator needs to be dressed to ensure the correct noninteracting behavior\cite{Kotliar_1986},
\begin{subequations}\label{dressed-SSpin}
\begin{eqnarray}
\tilde{S}^{+}_{is\sigma} &=& P^{+}_{is\sigma}S^{+}_{is\sigma}P^{-}_{is\sigma}\;, \\
P^{\pm}_{is\sigma} &=& \frac{1}{\sqrt{1/2 \pm S^{z}_{is\sigma}}}\;,
\end{eqnarray}
\end{subequations}
which can be further linearized as:
\begin{equation}
\tilde{S}^{+}_{is\sigma} \approx \tilde{z}^\dagger_{is\sigma}
+ \frac{\langle\tilde{z}^\dagger_{is\sigma}\rangle \langle S^z_{is\sigma} \rangle \Delta S^z_{is\sigma}}
{( \tfrac{1}{2} )^2 - \langle S^z_{is\sigma} \rangle^2 }
\end{equation}
with $\Delta S^z_{is\sigma} = S^z_{is\sigma} - \langle S^z_{is\sigma} \rangle$
and $\tilde{z}^\dagger_{is\sigma} =  S^{+}_{is\sigma} /
\sqrt{ (\tfrac{1}{2})^2 - \langle S^z_{is\sigma} \rangle^2 }$.

In the cluster slave spin method\cite{WCLee17,Zeng21,Zeng22,Zeng25_1}, the slave spin Hamiltonian ${\rm H}_{MF}^{S}$ can be decomposed into
a series of repetitive slave spin clusters with the $n_{c}$ sites, the interaction between slave spins within the cluster is dealt with accurately
using the exact diagonalization, while the coupling between the cluster and its environment is treated approximately using the conventional mean-field
decoupling $AB\approx \langle A\rangle B + A\langle B\rangle - \langle A\rangle \langle B\rangle$. 
In the ground state with compensated collinear magnetism such as that shown in Fig.\ref{subcurrent-2site-cluster},
the local constraints~\eqref{constraint} are enforced roughly by the global Lagrange multipliers
$\lambda_{I\sigma}$ on sublattices $I=A$ and $B$, Hamiltonian~\eqref{HH-Model-SSpin} can be therefore recast as
\begin{subequations}\label{HAMILTONIAN}
\begin{align}
{\rm H}^f_{\text{MF}} =& -tZ\sum_{\langle i,j\rangle \sigma} ( f_{iA\sigma}^\dagger f_{jB\sigma}+ \textsc{h.c.}) \nonumber\\
+& \lambda \sum_{i I \bm{\alpha} \sigma} Z_{I\sigma} e^{i\tfrac{\pi}{2}\nu_{i_{I}i_{I}+\bm{\alpha}}}f_{iI\sigma}^\dagger f_{i+\bm{\alpha}I\sigma} \nonumber\\
+& \sum_{iI\sigma}(\mu+\tilde{\mu}_{I\sigma}-\lambda_{I\sigma})f_{iI\sigma}^\dagger f_{iI\sigma} \;, \label{FERMI-HAML}\\
{\rm H}_{MF}^{S} =& \frac{2N}{n_c}{\rm H}_{n_c\text{-site}}^{S}\;, \nonumber\\
{\rm H}_{n_c\text{-site}}^{S} =& H^\lambda_{n_c\text{-site}}+H^U_{n_c\text{-site}}+H^K_{n_c\text{-site}}
\;, \label{SSPIN-HAML-nc-SITE}
\end{align}
\end{subequations}
where
\begin{eqnarray}
 \tilde{\mu}_{I\sigma} &=& \frac{2 \langle \tilde{z}^\dagger_{A\sigma}\rangle\langle\tilde{z}_{B\sigma}\rangle \langle S_{I\sigma}^z \rangle
 (\epsilon_\sigma^{\bm{\delta}_1} + \epsilon_\sigma^{\bm{\delta}_2} + \epsilon_\sigma^{\bm{\delta}_3}) }{( \tfrac{1}{2} )^2 - \langle S_{I\sigma}^z \rangle^2 }\nonumber\\
&+& (-1)^{s}\frac{2 \langle \tilde{z}^\dagger_{I\sigma}\rangle\langle\tilde{z}_{I\sigma}\rangle \langle S_{I\sigma}^z \rangle
\sum_{\bm{\alpha}}\epsilon_{I\sigma}^{\bm{\alpha}} }{( \tfrac{1}{2} )^2 - \langle S_{I\sigma}^z \rangle^2 }
\end{eqnarray}
with $s=0$, 1 for sublattice $I=$ $A$ and $B$, respectively; $\epsilon_\sigma^{\bm{\delta}}=-t\langle f_{iA\sigma}^{\dagger}f_{i+\bm{\delta}B\sigma} \rangle$, $\epsilon_{I\sigma}^{\bm{\alpha}}=i\lambda\nu_{i_{I}i_{I}+\bm{\alpha}}\langle f_{iI\sigma}^{\dagger}f_{i+\bm{\alpha}I\sigma} \rangle$ with $\bm{\delta}_{i=1,2,3}=C_{3}^{i-1}\tfrac{1}{\sqrt{3}}(1,0)^{T}$ and $\bm{\alpha}=\pm\bm{a}_{1}, \pm\bm{a}_{2}, \pm(\bm{a}_{1}+\bm{a}_{2})$ representing
respectively the NN and NNN vectors in which the real-space unit vectors read $\bm{a}_{1}=(-\tfrac{\sqrt{3}}{2},\tfrac{1}{2})^{T}$
and $\bm{a}_{2}=(\tfrac{\sqrt{3}}{2},\tfrac{1}{2})^{T}$; $N$ is the number of unit cells in the hexagon lattice; ${\rm H}_{n_c\text{-site}}^{S}$ denotes
the cluster slave spin Hamiltonian expanded on the basis of the cluster with $n_c$ sites\cite{Zeng21,Zeng22},
and $H^{\lambda,K,U}_{n_c\text{-site}}$ represents its components proportional to $\lambda_{I\sigma}$, the kinetic energy term, and the onsite Coulomb interaction term, respectively. In particular, unlike the single-site approximation,
the dressed slave spin operator in Eq.\eqref{dressed-SSpin} can not give rise to the correct noninteracting
behavior, i.e., $Z=1$ at $U=0$\cite{Zeng21,Zeng22}. Then the parameters $\alpha_{1/2/3}$ have been introduced to ensure the correct noninteracting behavior,
thus the quasiparticle residues are written as $Z=\alpha_1\langle \tilde{z}^\dagger_{A\sigma}\rangle\langle\tilde{z}_{B\sigma}\rangle$,
$Z_{I\sigma}=\alpha_{2/3}\langle \tilde{z}^\dagger_{I\sigma}\rangle \langle\tilde{z}_{I\sigma}\rangle$\cite{Zeng25_1}. 

Upon the Fourier transformation, the fermionic component of Hamiltonian~\eqref{HAMILTONIAN} can be written in momentum space as
\begin{eqnarray}\label{FERMI-HAML-k}
{\rm H}^f_{\text{MF}} &=& \sum_{\bm{k}\sigma}\Psi_{\bm{k}\sigma}^{\dagger} H_{\bm{k}\sigma}\Psi_{\bm{k}\sigma}
\end{eqnarray}
with
\begin{equation*}
H_{\bm{k}\sigma} = \left(\begin{array}{cc}
\epsilon_{\bm{k}A\sigma}, & -tZ\gamma_{\bm{k}} \\
-tZ\gamma_{\bm{k}}^{\ast}, & \epsilon_{\bm{k}B\sigma}
\end{array} \right)\;, \;\;  \Psi_{\bm{k}\sigma} = (f_{A\bm{k}\sigma},f_{B\bm{k}\sigma})^{T} \;,
\end{equation*}
in which
\begin{align*}
\varepsilon_{\bm{k}I\sigma} &= \mu+\tilde{\mu}_{I\sigma}-\lambda_{I\sigma} +(-1)^{s}Z_{I\sigma}2\lambda\gamma_{2\bm{k}}\;, \\
\gamma_{\bm{k}} &= \sum_{\bm{\delta}}e^{i\bm{k}\cdot\bm{\delta}}\;, \;\;\;
2\gamma_{2\bm{k}} = \sum_{\bm{\alpha}}e^{i\tfrac{\pi}{2}\nu_{i_{A}i_{A}+\bm{\alpha}}}e^{i\bm{k}\cdot\bm{\alpha}}\;.
\end{align*}
From the Hamiltonian~\eqref{FERMI-HAML-k}, the spinon Green's function can be directly derived as,
\begin{align}\label{spinon-GFunction}
&\tilde{G}_{f\sigma}(\bm{k},i\omega) = [ i\omega - H_{\bm{k}\sigma} ]^{-1} \nonumber\\
&= \frac{\left[ \begin{array}{ll}
i\omega-\epsilon_{\bm{k}B\sigma},& -tZ\gamma_{\bm{k}} \\
-tZ\gamma_{\bm{k}}^{\ast},& i\omega-\epsilon_{\bm{k}A\sigma}
\end{array}\right]}{(i\omega-\epsilon_{\bm{k}A\sigma})(i\omega-\epsilon_{\bm{k}B\sigma})-|tZ\gamma_{\bm{k}}|^2}
\end{align}
with the fermionic Matsubara frequency $i\omega$. 

The diagonalization of the fermionic spinon Hamiltonian~\eqref{FERMI-HAML-k} gives rise to the following spinon energy dispersion,
\begin{eqnarray}\label{spinon-disp}
E_{\bm{k}\sigma}^{(\pm)}&=&\mu_{\rm eff}+\lambda(Z_{A\sigma}-Z_{B\sigma})\gamma_{2\bm{k}} \nonumber\\
&\pm& \sqrt{ [\Delta_{\sigma}+\lambda(Z_{A\sigma}+Z_{B\sigma})\gamma_{2\bm{k}} ]^2 + |tZ\gamma_{\bm{k}}|^2 } \;,
\end{eqnarray}
where
\begin{align*}
\mu_{\rm eff} &= \mu + \frac{1}{2}( \tilde{\mu}_{A\sigma}-\lambda_{A\sigma}+\tilde{\mu}_{B\sigma}-\lambda_{B\sigma} )\;, \\
\Delta_{\sigma} &= \frac{1}{2}( \tilde{\mu}_{A\sigma}-\lambda_{A\sigma}-\tilde{\mu}_{B\sigma}+\lambda_{B\sigma} )\;.
\end{align*}
In the presence of compensated collinear magnetism, there exists a relation $\tilde{\mu}_{B\sigma}=\tilde{\mu}_{A-\sigma}$
and $\lambda_{B\sigma}=\lambda_{A-\sigma}$, indicating that
\begin{equation}
\Delta_{\sigma} = \frac{1}{2}( \tilde{\mu}_{A\sigma}-\lambda_{A\sigma}-\tilde{\mu}_{A-\sigma}+\lambda_{A-\sigma} )\;,
\end{equation}
therefore we have $\Delta_{\sigma}=-\Delta_{-\sigma}$, which reflects the breaking symmetry $[C_2||\bm{\tau}]$ by the Haldane hopping with $C_2$ and $\bm{\tau}$ respectively
denoting a rotation of $180^{\circ}$ around the axis perpendicular to spins and the minimal vector connecting opposite-spin sublattices. Simultaneously,
the function $\gamma_{2\bm{k}}$ coming from the Haldane hopping that is odd under the inversion breaks the symmetry $[\bar{C}_2||T]$ with $T$ only acting in real space.
However, the above breaking symmetries combined give rise to the spinon energy gap that is odd under the transformation $[C_2||\bar{E}]$,
\begin{equation}\label{spinon-gap}
\Delta_{\bm{k}\sigma} = \Delta_{\sigma}+\lambda(Z_{A\sigma}+Z_{B\sigma})\gamma_{2\bm{k}},
\end{equation}
which thus accounts for the $f$-wave odd-parity ALM appearing in HH model with the symmetry $[C_2||\bar{E}]$.
In the vicinity of Dirac points $\bm{K}$ and $\bm{K}'$ with $\bm{k}\approx \bm{K}^{(')}+\bm{q}$, the spinon energy gap can be further reduced as,
\begin{equation}\label{spinon-gap-DP}
\Delta_{\bm{q}\sigma} \approx \Delta_{\sigma}+\lambda(Z_{A\sigma}+Z_{B\sigma})(-s\frac{3\sqrt{3}}{2}) + 0(q^2)
\end{equation}
with $s=\pm1$ for the Dirac point $\bm{K}$ and $\bm{K}'$, respectively. Therefore, in paramagnetic states with the vanishing AFM energy gap $\Delta_{\sigma}$,
the Haldane hopping opens opposite energy gaps proportional to $\lambda$ at the Dirac point $\bm{K}$ and $\bm{K}'$, which drives the system
into the CI phase.

Furthermore, the slave-spin component of Hamiltonian~\eqref{HAMILTONIAN} can be approximated as
within the 2-site cluster slave-spin approximation with $n_c = 2$[cf. Fig.\ref{subcurrent-2site-cluster}(b)]:
\begin{subequations}\label{SSPIN-H}
\begin{align}
{\rm H}^\lambda_{2\text{-site}} =& \sum_{i_c=1,\sigma}^{2}\lambda_{I\sigma}( S_{i_c\sigma}^{z}+\frac{1}{2} )\;, \\
{\rm H}^U_{2\text{-site}} =& \sum_{i_c=1}^{2}U(S_{i_c\sigma}^{z}+\tfrac{1}{2})(S_{i_c-\sigma}^{z}
+\tfrac{1}{2})\;, \\
{\rm H}^K_{\text{2-site}} =& \sum_{\sigma} \Big\{ \epsilon_{\sigma}^{\bm{\delta}_1} ( \tilde{z}_{A\sigma}^\dagger
\tilde{z}_{B\sigma} +\tilde{z}_{B\sigma}^\dagger \tilde{z}_{A\sigma} )\nonumber\\
&+ (\epsilon_{\sigma}^{\bm{\delta}_2} +  \epsilon_{\sigma}^{\bm{\delta}_3} )
\big[ \tilde{z}_{A\sigma}^\dagger \langle \tilde{z}_{B\sigma} \rangle
+ \tilde{z}_{B\sigma}^\dagger \langle \tilde{z}_{A\sigma} \rangle + \textsc{h.c.} \big] \nonumber\\
&+ \sum_{I}(-1)^{s}[ \tilde{z}_{I\sigma}^\dagger \langle \tilde{z}_{I\sigma} \rangle + \textsc{h.c.} ]
\sum_{\bm{\alpha}}\epsilon_{I\sigma}^{\bm{\alpha}} \Big\}\;,
\end{align}
\end{subequations}
where $s=$ 0, 1 holds for sublattice $I=$ $A$ and $B$, respectively; In this work, the $n_c$-site cluster slave spin Hamiltonian
is solved via the numerical method of exact diagonalization. Now we show that the constraints \eqref{local-constraint} are ensured roughly
at two sublattices only if the slave spin cluster consists of $n_c/2$ unit cells,
\begin{align}
&\partial_{\lambda_{I\sigma}}\langle {\rm H}\rangle \approx \partial_{\lambda_{I\sigma}}\langle {\rm H}^f_{\text{MF}}
+ \frac{2N}{n_c}{\rm H}_{n_c\text{-site}}^{S}\rangle \nonumber\\
&\approx -\sum_{iI'\sigma'} \langle f_{iI'\sigma'}^{\dagger}f_{iI'\sigma'}\rangle \delta_{I\sigma,I'\sigma'} \nonumber\\
&+ \frac{2N}{n_c}\frac{n_c}{2}\sum_{I'\sigma'}( \langle S_{I'\sigma'}^{z}\rangle+\frac{1}{2} ) \delta_{I\sigma,I'\sigma'}\nonumber\\
&= -\sum_{i} \langle f_{iI\sigma}^{\dagger}f_{iI\sigma}\rangle + N( \langle S_{I\sigma}^{z}\rangle+\frac{1}{2} ) = 0\;,
\end{align}
which directly gives rise to that
\begin{equation}
\langle S_{I\sigma}^{z}\rangle = \frac{1}{N}\sum_{i} \langle f_{iI\sigma}^{\dagger}f_{iI\sigma}\rangle-\frac{1}{2}\;.
\end{equation}
Therefore, we conclude that the constraints \eqref{local-constraint} are indeed ensured at the mean-field level. 
We note that the quantities $\langle \tilde{z}_{A/B\sigma} \rangle$, $\langle S_{I\sigma}^z \rangle$, $\epsilon_{\sigma}^{\bm{\delta}}$,
$\epsilon_{I\sigma}^{\bm{\alpha}}$, $\mu$, and Lagrange multipliers $\lambda_{I\sigma}$ are self-consistently determined for the approximate
Hamiltonian~\eqref{HAMILTONIAN}.

On the basis of the slave spin representation, the electron Green's function can be approximated as using the mean-field decoupling,
\begin{eqnarray}
G_{II'\sigma}(i-j,\tau) &=& - \langle T_{\tau}C_{iI\sigma}(\tau)C_{jI'\sigma}^{\dagger} \rangle  \nonumber\\
&\approx& -\langle T_{\tau}\tilde{z}_{iI\sigma}(\tau)\tilde{z}_{jI'\sigma}^{\dagger} \rangle
\langle T_{\tau}f_{iI\sigma}(\tau)f_{jI'\sigma}^{\dagger} \rangle \;, \nonumber\\
\end{eqnarray}
which can be further simplified in the itinerant state with the condensation of the composite bosonic field $\tilde{z}^\dagger_{is\sigma}$,
\begin{eqnarray}
G_{II'\sigma}(i-j,\tau) &\approx& -\alpha_{1/2/3}\langle \tilde{z}_{I\sigma}\rangle \langle \tilde{z}_{I'\sigma}^{\dagger}\rangle
\langle T_{\tau}f_{iI\sigma}(\tau)f_{jI'\sigma}^{\dagger} \rangle \nonumber\\
&=& \alpha_{1/2/3}\langle \tilde{z}_{I\sigma}\rangle \langle \tilde{z}_{I'\sigma}^{\dagger}\rangle G_{II'\sigma}^{f}(i-j,\tau)\;.
\end{eqnarray}
Therefore, by Fourier transforming the spinon Green's function $G_{II'\sigma}^{f}(i-j,\tau)$ into momentum space, the electron Green's function in momentum space
can be directly obtained as $G_{II'\sigma}(\bm{k},i\omega) = \alpha_{1/2/3}\langle \tilde{z}_{I\sigma}\rangle \langle \tilde{z}_{I'\sigma}^{\dagger}\rangle G_{II'\sigma}^{f}(\bm{k},i\omega)$ with $\alpha_{1/2/3}\langle \tilde{z}_{I\sigma}\rangle \langle \tilde{z}_{I'\sigma}^{\dagger}\rangle$ being the quasiparticle
residue. Then using the fermionic spinon Green's function \eqref{spinon-GFunction}, the electron Green's function
$\tilde{G}_{\sigma}(\bm{k},i\omega)=\int_{0}^{\beta}d\tau e^{i\omega\tau}\big[-\langle T_{\tau}\Psi_{\bm{k}\sigma}(\tau)\Psi_{\bm{k}\sigma}^{\dagger}(0)\rangle\big]$
with $\Psi_{\bm{k}\sigma}^{\dagger}=(C_{\bm{k}A\sigma}^{\dagger},C_{\bm{k}B\sigma}^{\dagger})$ can be directly expressed as
\begin{equation}\label{EL-GFunction}
\tilde{G}_{\sigma}(\bm{k},i\omega) = \frac{\left[
\begin{array}{ll}
Z_{A\sigma}(i\omega-\epsilon_{\bm{k}B\sigma}),& -tZ^2\gamma_{\bm{k}} \\
-tZ^2\gamma_{\bm{k}}^{\ast},& Z_{B\sigma}(i\omega-\epsilon_{\bm{k}A\sigma})
\end{array}\right]}{(i\omega-\epsilon_{\bm{k}A\sigma})(i\omega-\epsilon_{\bm{k}B\sigma})-|tZ\gamma_{\bm{k}}|^2} \;.
\end{equation}

\subsection{The optical conductivity}

Now we turn to deriving the optical conductivity of HH model using the electron Green's function from the cluster slave spin method.
The linear response theory gives rise to the optical conductivity in terms of the Kubo formula\cite{Mahan90}
\begin{equation}\label{conductivity-1}
\sigma_{ij}(\Omega,T) = -\frac{{\rm Im}\Pi_{ij}(\Omega)}{\Omega},
\end{equation}
where $\Pi_{ij}(\Omega)$ is the retarded electron current-current correlation function with $i$, $j$ being $x$ and $y$, respectively.
Its counterpart in the Matsbara representation can be directly calculated as,
\begin{equation}\label{corP-1}
\Pi_{ij}(i\Omega_{m})=-\frac{1}{V}\int_{0}^{\beta}d\tau e^{i\Omega_{m}\tau}\langle T_{\tau}J_{i}(\tau)J_{j}(0)\rangle
\end{equation}
with the volume $V$, the bosonic Matsubara frequency $\Omega_{m}=2\pi m/\beta$, and the current density of electrons $\bm{J}$.
This electric current density can be obtained in terms of the electron polarization operator,
which is a summation over all electrons and their positions\cite{Mahan90,Zeng23}, i.e., ${\bf P}=\sum_{l\sigma}{\bf R}_{l}C^{\dag}_{l\sigma}C_{l\sigma}$.
The electric current density can be obtained by evaluating the time derivative of the polarization operator using the Heisenberg's equation,
\begin{eqnarray}\label{current-density-1}
{\bf J} &=& \frac{ie}{\hbar}[{\bf P},{\rm H}]
= \frac{e}{\hbar} \big\{ -it\sum_{\bm{k}\bm{\delta}\sigma}\bm{\delta}[-C_{\bm{k}A\sigma}^{\dagger}C_{\bm{k}B\sigma}e^{i\bm{k}\cdot\bm{\delta}} + \text{H.C.}] \nonumber\\
&+& 2\lambda\sum_{\bm{k}\bm{\alpha}s\sigma}(-1)^{s}\bm{\alpha}\cos(\bm{k}\cdot\bm{\alpha})C_{\bm{k}s\sigma}^{\dagger}C_{\bm{k}s\sigma} \big\}\;.
\end{eqnarray}

The components of electric current in Eq.\eqref{current-density-1} along the $x$ and $y$ direction can be calculated as
$J_{x/y} = \tfrac{e}{\hbar}\sum_{\bm{k}\sigma}\Psi_{\bm{k}\sigma}^{\dagger}\tilde{v}_{\bm{k}}^{(x/y)}\Psi_{\bm{k}\sigma}$ with
\begin{subequations}
\begin{eqnarray}
\tilde{v}_{\bm{k}}^{(x)} &=& \left[\begin{array}{l}
-2\sqrt{3}\lambda\sin\frac{\sqrt{3}}{2}k_x\sin\frac{k_y}{2},\\
\frac{it}{\sqrt{3}}(e^{\tfrac{ik_x}{\sqrt{3}}}-e^{\tfrac{-ik_x}{2\sqrt{3}}}\cos\frac{k_y}{2}); \\
\frac{-it}{\sqrt{3}}(e^{\tfrac{-ik_x}{\sqrt{3}}}-e^{\tfrac{ik_x}{2\sqrt{3}}}\cos\frac{k_y}{2}),\\
2\sqrt{3}\lambda\sin\frac{\sqrt{3}}{2}k_x\sin\frac{k_y}{2}
\end{array} \right]\;, \\
\tilde{v}_{\bm{k}}^{(y)} &=& \left[\begin{array}{l}
2\lambda(\cos\frac{\sqrt{3}}{2}k_x\cos\frac{k_y}{2}-\cos k_y),\\
-te^{\tfrac{-ik_x}{2\sqrt{3}}}\sin\frac{k_y}{2}; \\
-te^{\tfrac{ik_x}{2\sqrt{3}}}\sin\frac{k_y}{2},\\
-2\lambda(\cos\frac{\sqrt{3}}{2}k_x\cos\frac{k_y}{2}-\cos k_y)
\end{array} \right]\;.
\end{eqnarray}
\end{subequations}
The current-current correlation function in Eq.\eqref{corP-1} is therefore evaluated as
\begin{eqnarray}\label{corP-2}
&&\Pi_{ij}(i\Omega_{m}) \nonumber\\
&&= \frac{e^2}{N\hbar^2\beta}\sum_{\bm{k}i\omega \sigma}{\rm Tr}
[\tilde{v}_{\bm{k}}^{(i)}\tilde{G}_{\sigma}(\bm{k},i\omega_n+i\Omega_m)\tilde{v}_{\bm{k}}^{(j)}\tilde{G}_{\sigma}(\bm{k},i\omega_n)],\nonumber\\
\end{eqnarray}
then the retarded current-current correlation function is reached via the analytical continuation, i.e., $i\Omega_{m}\to \Omega+i\Gamma$ where
$\Gamma=0.01$ is used in this work.

\section{Quantitative characteristics}\label{Results}

\subsection{Electronic properties}\label{Results-EDisp}

\begin{figure}[t]
\includegraphics[scale=0.35]{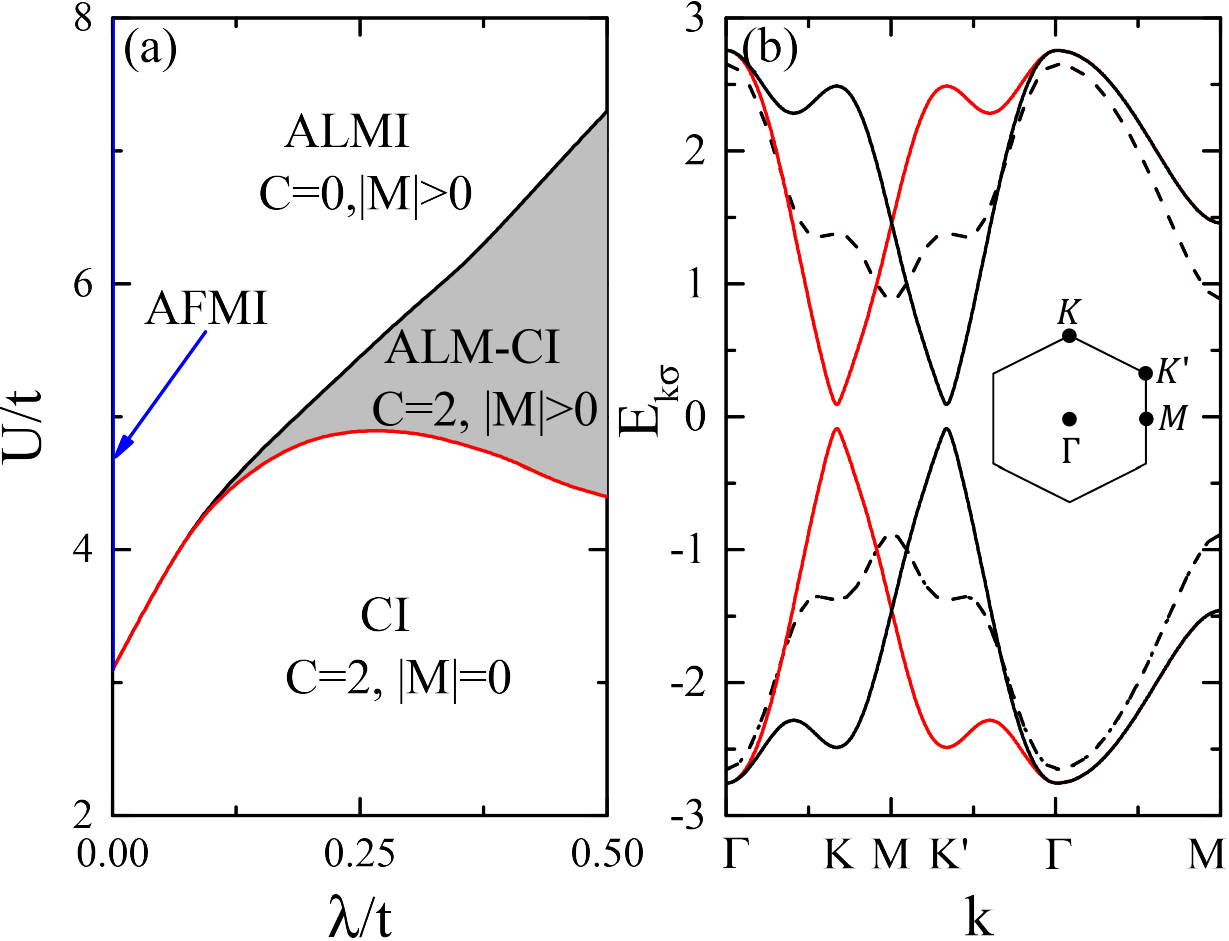}
\caption{(Color online) (a)The phase diagram of HH model as a function of Haldane hopping $\lambda$ and onsite electron
Coulomb repulsion $U$, which consists of four regimes: AFMI at $\lambda=0$ for $U>U_{\rm AFM}\approx 3.1$,
CI with the Thouless-Kohmoto-Nightingale-Nijs number $C=2$ at small $U's$,
the odd-parity ALM-CI phase with $C=2$ and $|M|>0$ in the intermediate interaction regime, and odd-parity ALMI
at large interaction strengths[From Ref. \onlinecite{Zeng25_1}].
(b)The momentum dependence of the spinon energy dispersion $E_{\bm{k}\sigma}$ in Eq.\eqref{spinon-disp}
along the zigzag direction in the reduced first Brillouin zone(FBZ), illustrated by the inset where two inequivalent Dirac points $\bm{K}$ and $\bm{K}'$
and their middle point $\bm{M}$ are marked, at $\lambda=0.3$ as well as $U=$4.8(black dashed) and 5.5 for up(red solid) and down(black solid) spinons.
\label{PD-Spinon-Energy}}
\end{figure}
\begin{figure*}
\includegraphics[scale=0.5]{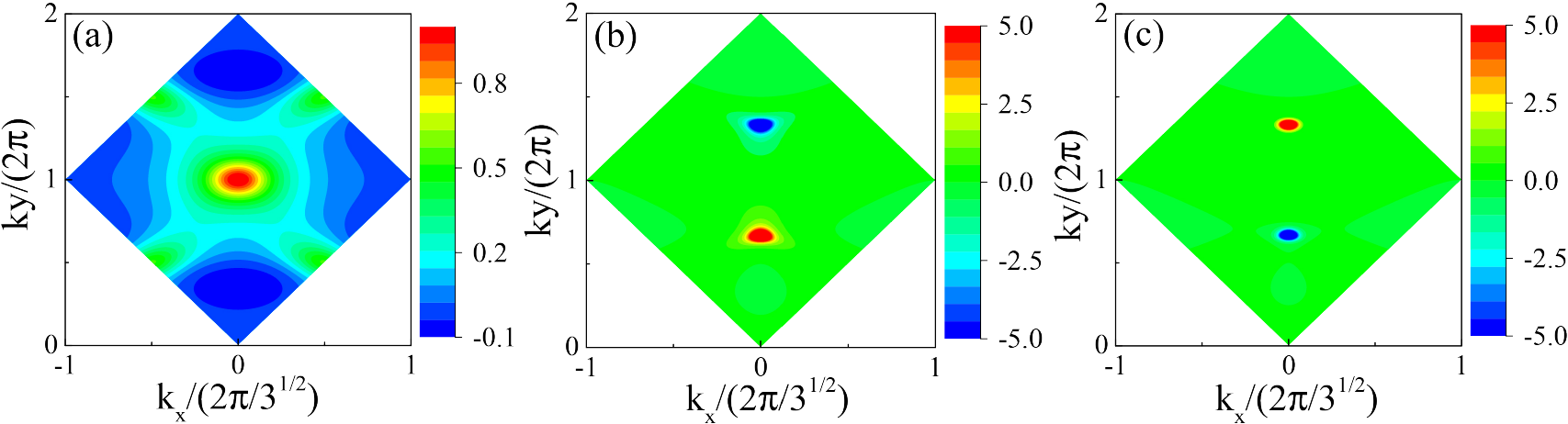}
\caption{(Color online)The spin resolved spinon Berry curvature $B_{\sigma}(\bm{k})$ as a function of momentum at $\lambda=0.3$
with (a)$U=4.8$(CI), (b)$5.5$(ALM-CI), and (c)$6.1$(ALMI) in the FBZ.
\label{BCurvature} }
\end{figure*}
To facilitate the discussion about the topological physics of HH model, we in Fig.\ref{PD-Spinon-Energy}(a) replot its phase diagram
as a function of Haldane hopping $\lambda$ and the onsite electron Coulomb repulsion $U$. After self-consistently
determining the mean-field parameters $\langle \tilde{z}_{I\sigma} \rangle$, $\langle S_{I\sigma}^z \rangle$, $\epsilon_{\sigma}^{\bm{\delta}}$,
$\epsilon_{I\sigma}^{\bm{\alpha}}$, and Lagrange multipliers $\lambda_{I\sigma}$ with $I=$ A and B, the phase diagram is determined
via the staggered magnetization $M$ and the Chern number $C$ which can be calculated as
\begin{equation}\label{MF-PARAM}
M = \frac{1}{4N}\sum_{is}(-1)^s[\langle C_{is\uparrow}^{\dagger}C_{is\uparrow} \rangle -\langle C_{is\downarrow}^{\dagger}C_{is\downarrow} \rangle]\;,
\end{equation}
with $s=$0, 1 for sublattice A and B, respectively, and
\begin{equation}
C = \frac{1}{2\pi}\sum_{n\sigma}\int_{\rm BZ} d^{2}\bm{k}\Omega_{\sigma}^{(n)}(\bm{k})n_{\rm F}[E^{(n)}(\bm{k})]
\end{equation}
with the Berry curvature evaluated as\cite{Qiao10}
\begin{eqnarray}\label{BCurve}
\Omega_{\sigma}^{(n)}(\bm{k})&=& -2\sum_{n'\neq n}{\rm Im}\{ \frac{\langle U_{\bm{k}\sigma}^{(n)}|\partial_{k_x}{\rm H}_{\bm{k}\sigma}|U_{\bm{k}\sigma}^{(n')}\rangle
}{(E_{\bm{k}\sigma}^{(n)}-E_{\bm{k}\sigma}^{(n')})^2}\nonumber\\
&\times& \langle U_{\bm{k}\sigma}^{(n')}|\partial_{k_y}{\rm H}_{\bm{k}\sigma}|U_{\bm{k}\sigma}^{(n)}\rangle \}\;,
\end{eqnarray}
and the fermionic distribution
\begin{equation}
n_{\rm F}[E^{(n)}(\bm{k})] = \frac{1}{2}\big[ 1-\tanh\frac{\beta E^{(n)}(\bm{k})}{2} \big]\;,
\end{equation}
where $\beta = 1/T$ is the inverse temperature. Please note that the staggered magnetization $M$ itself is not sufficient for identifying the odd-parity ALM
with the compensated collinear magnetism, while the odd-parity spin splitting in momentum space subjected to the symmetry $[C_2||\bar{E}]$ is an indispensable evidence\cite{Zeng25_1}[See also Fig.\ref{PD-Spinon-Energy}(b)]. 
As shown by the intermediate interaction regime plotted in grey, there appears the topologically nontrivial ALM-CI phase at large Haldane hoppings,
where the $f$-wave odd-parity ALM having the nonzero $M$ coexists with the Chern insulator having $C=2$.
In particular, this topologically nontrivial ALM-CI phase with $C=2$ at around moderate interaction strengths has also been found
using the dynamical cluster approximation\cite{ifmmode16} and the unbiased mean-field theory\cite{Zheng15,Arun16}, except
that the ALM-CI phase has been inappropriately classified as the antiferromagnetic CI phase because the authors have neglected studying the
spin splitting in momentum space and the spin group symmetry of HH model. In addition, the severe sign problem restricts the application of the
QMC method to the HH model\cite{Loh90,Troyer05,Zheng15,Arun16,Wu16,Vanhala16,ifmmode16,Mertz19,He24,Mai23} 
Furthermore, in this regime, the spinon energy dispersion at $\lambda=0.3$ is systematically investigated,
which is equivalent to the electron counterpart before the occurrence of metal-insulator Mott
transition with the condensed composite bosonic field, i.e., $\langle \tilde{z}_{I\sigma}\rangle>0$. As shown by the black dashed line in Fig.\ref{PD-Spinon-Energy}(b),
the topological energy gap coming from large Haldane hoppings is greater than the energy separation centered around the midpoint $\bm{M}$ between the Dirac points
$\bm{K}$ and $\bm{K}'$, which moves the energy extrema from the Dirac points to their midpoint $\bm{M}$. Then because the Berry curvature in two-band systems
is inversely proportional to the square of the energy difference, the Berry curvature at $\lambda=0.3$ and $U<U_{\rm ALM}$ is therefore expected to be maximized
around the momentum $\bm{M}$ with $U_{\rm ALM}$ the critical interaction strength responsible for the transition into the ALM-CI phase.
However, as the staggered magnetization together with the AFM energy gap $\Delta_{\sigma}$ set in, the up-spin spinon energy gap[cf. Eq.\eqref{spinon-gap-DP}]
at momentum $\bm{K}$ and $\bm{K}'$ is strongly reduced and enhanced, respectively, while the opposite applies to the down-spin counterpart,
which is expected to cause pronounced impacts on the topological properties of HH model.

In Fig.\ref{BCurvature}, we further investigate the spin-resolved spinon Berry curvature $B_{\sigma}(\bm{k}) = \Omega_{\sigma}(\bm{k})\times \sigma$
as a function of momentum at $\lambda=0.3$ with (a)$U=4.8$, (b)$5.5$, and (c)$6.1$ in the FBZ, where $\sigma=\pm1$ is introduced
to project the Berry curvature into the up and down spin channel, respectively. As shown in Fig.\ref{BCurvature}(a),
the positive Berry curvature in the CI phase at $\lambda=0.3$ and $U=4.8$ is spin-degenerate
and accommodated at around the midpoint $\bm{M}$ between the Dirac points $\bm{K}$ and $\bm{K}'$.
However, as shown in Fig.\ref{BCurvature}(b), when the transition into the odd-parity ALM-CI phase occurs with the nonzero AFM energy gap $\Delta_{\sigma}$,
the Berry curvature for up- and down-spin spinons is centered around $\bm{K}$ and $\bm{K}'$, respectively,
which is consistent with the alternating characteristics of spin polarizations enforced by the symmetry $[C_2||\bar{E}]$.
While the alternating polarization for up- and down-spin Berry curvatures does not lead to different Chern numbers,
because the symmetry $[C_2||\bar{E}]$ ensures that the summation of the Berry curvature in the FBZ always gives rise to the same result,
i.e., $C=1$ for up and down spinons.
This distinction is conceptually important: unlike the previously studied correlation-driven HH transitions that are usually diagnosed by a change of the total Chern invariant or by gap closing\cite{Zheng15,Vanhala16,Mertz19,He24}, the present CI-to-ALM-CI transition keeps the quantized Hall topology intact but reconstructs where the Berry curvature is concentrated in momentum space.
In this sense, the odd-parity ALM order acts as a momentum-space topology reshuffler rather than as a mechanism that simply destroys the CI phase.
As shown in Fig.\ref{BCurvature}(c), when the HH model evolves from the ALM-CI phase into the ALMI phase,
the sign of the Berry curvature changes, and its strength around $\bm{K}$ and $\bm{K}'$ respectively for up- and down-spin spinons
is significantly weakened. As a result, the negative Berry curvature polarized in the vicinity of the Dirac points offsets the positive contribution distributed around $\bm{M}$.
Therefore, the system at $\lambda=0.3$ and $U=6.1$ still exhibits the alternating spin polarization
enforced by the symmetry $[C_2||\bar{E}]$, while becomes topologically trivial; we in this sense identify the system in this phase as an ALM insulator.

\begin{figure}[t]
\includegraphics[scale=0.42]{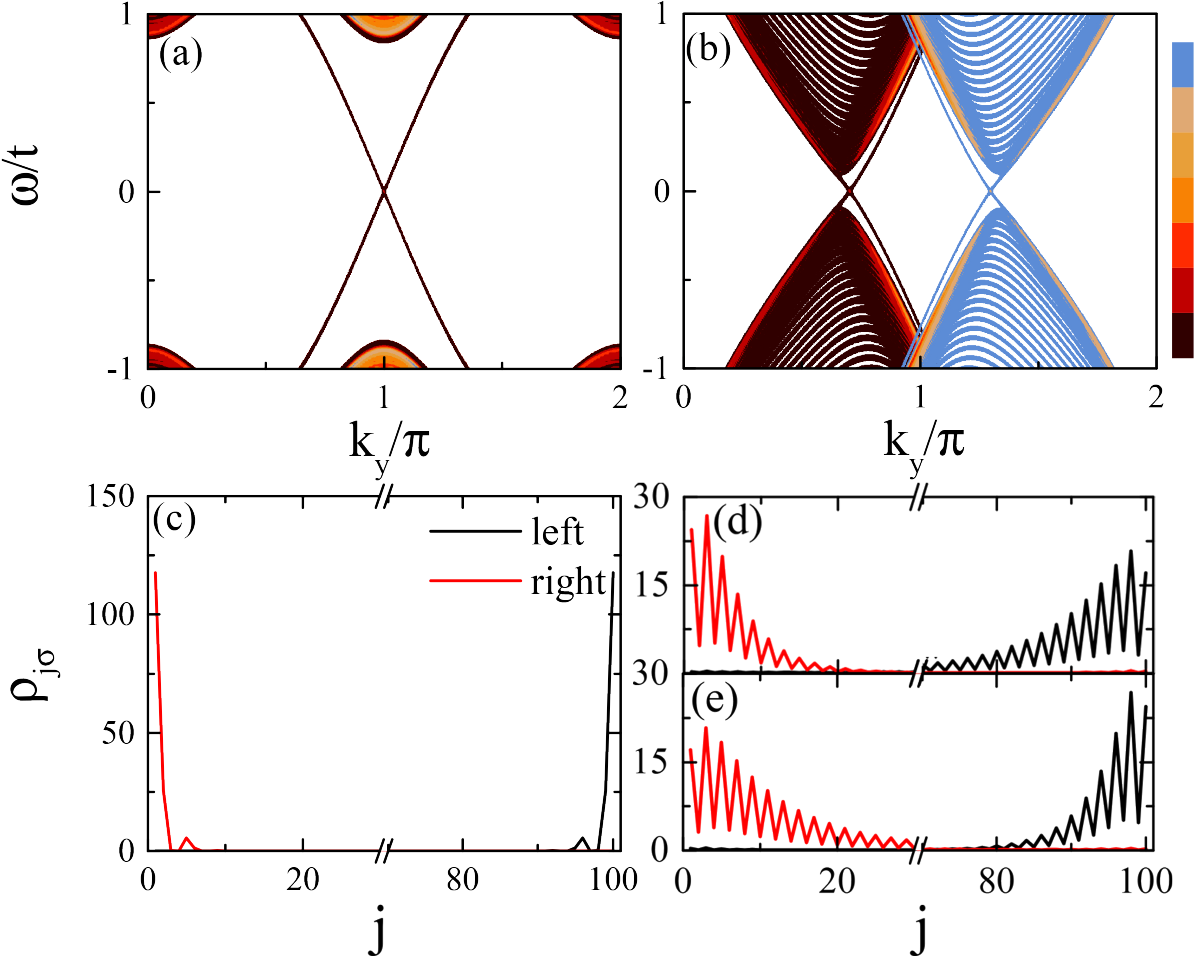}
\caption{(Color online)The zigzag energy dispersion as a function of momentum $k_y$ mapped by the electron spectrum function
$A_{\sigma}(k_y,\omega)$ at $\lambda=0.3$ with (a)$U=4.8$(CI) and (b)$5.5$(ALM-CI), respectively. (c)the electronic LDOS of the edge mode as a
function of site location $j$ in the CI phase with $\lambda=0.3$ and $U=4.8$; (d)the up- and (e)down-spin electronic LDOS of the edge mode
in the ALM-CI phase with $\lambda=0.3$ and $U=5.5$. Here, the the color scale at the right-hand side in the first row measures the spectral weight
for up-spin electrons; the legend "left" and "right" in subfigure (c) represent the left and right edge mode in the first row, respectively.
\label{Zigzag-Edgestate} }
\end{figure}
The above results show that the Berry curvature changes qualitatively across the transition from the conventional CI phase into the ALM-CI phase.
It is therefore natural to ask how the edge states evolve simultaneously, since they are an indispensable component of the topological response.
Based on the cluster slave-spin solution for the HH model [cf. Subsection~\ref{HH-Cluster-SSpin}], the electron Green's function of the ribbon with periodicity
in the zigzag direction can be directly obtained.
The electron spectrum function is then obtained as $A_{\sigma}(k_y,\omega)=-2\sigma{\rm Im}Tr[\tilde{G}_{\sigma}(k_y,\omega)]$,
where $\sigma=\pm1$ represents respectively the up and down spin polarization; $\tilde{G}_{\sigma}^{(ij)}(k_y,\omega)=Z_{ij}\tilde{G}_{f\sigma}^{(ij)}(k_y,\omega)$
with the renormalization factor $Z_{ij}\approx Z_{ss'}$ and the spinon Green's function calculated as $\tilde{G}_{f\sigma}(k_y,\omega)^{-1}=\omega-\tilde{\rm H}_{f}(k_y)$.
As shown in Fig.\ref{Zigzag-Edgestate}(a), the zigzag spectrum function $A_{\sigma}(k_y,\omega)$ in the CI phase is degenerate for up- and down-spin electrons,
and is symmetric with respect to $k_y=\pi$ and $\omega=0$, as required by the inversion symmetry, i.e., $E_{-\pi-q_y\sigma}=E_{\pi-q_y\sigma}=E_{\pi+q_y\sigma}$,
and the particle-hole symmetry, respectively. In addition, edge states in the CI phase exhibit the chiral symmetry,
which is also reflected in its local density of states(LDOS) distribution, as demonstrated in Fig.\ref{Zigzag-Edgestate}(c).
However, as shown in Fig.\ref{Zigzag-Edgestate}(b), the breaking inversion symmetry in the ALM-CI phase enables that the crossing of edge modes
deviates from the inversion-invariant momentum $k_y=\pi$, while the energy dispersion of the edge modes is subjected to the relation
$E_{\pi-q_y\sigma}=E_{\pi+q_y-\sigma}$, as required by the symmetry $[C_2||\bar{E}]$.
In particular, as shown in Figs.\ref{Zigzag-Edgestate}(c) and (d) where the LDOS of edge modes has been studied,
the zigzag ribbon of HH model in the ALM-CI phase exhibits the nonsymmetric chiral edge states, which is one of the main findings in this work.
Combined with the Berry-curvature reconstruction discussed above, this special edge mode indicates that the odd-parity ALM order does not merely split the pre-existing topological edge spectrum. Instead, it changes the symmetry constraint on the boundary problem itself, converting the conventional inversion-symmetric zigzag crossing into a spin-valley-locked edge structure.

\begin{figure}[!t]
\includegraphics[scale=0.42]{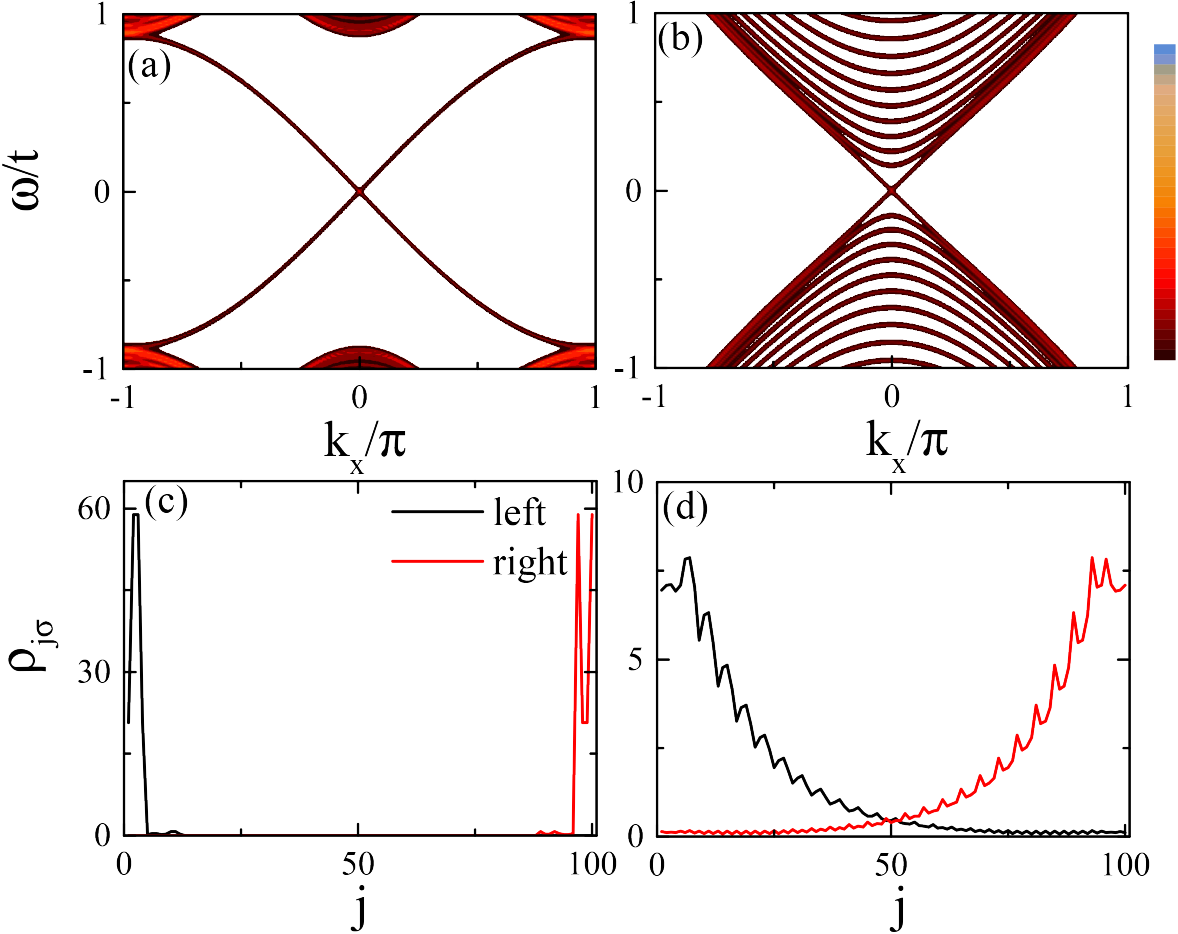}
\caption{(Color online)The armchair energy dispersion as a function of momentum $k_x$ mapped by the electron spectrum function
$A_{\sigma}(k_x,\omega)$ at $\lambda=0.3$ with (a)$U=4.8$(CI) and (b)$5.5$(ALM-CI). The corresponding electron LDOS of the edge modes
as a function of site location $j$ are respectively plotted in subfigure (c) and (d).
Here, the legend "left" and "right" represent the left and right edge mode in the first row, respectively.
\label{Armchair-Edgestate} }
\end{figure}
We have just shown that the inversion-symmetry breaking in the ALM-CI phase shifts the crossing of the edge modes
in the zigzag ribbon from the inversion invariant momentum $k_y=\pi$, and eventually gives rise to the nonsymmetric chiral edge states.
It is therefore meaningful to further clarify the characteristics of the armchair edge states.
In Figs.\ref{Armchair-Edgestate}(a) and (b), we present the edge states of the ribbon periodical in the armchair direction for $\lambda=0.3$
with (a)$U=4.8$(CI) and (b)$5.5$(ALM-CI). The corresponding LDOS are shown in Figs.\ref{Armchair-Edgestate}(c) and (d).
The main characteristics of the armchair edge states can be summarized as follows:
(i)the edge-state energy dispersion is degenerate for up- and down-spin electrons, i.e., $[C_2||E]E_{k_x\sigma}=E_{k_x-\sigma}$;
(ii)both the edge modes in the CI and ALM-CI phase exhibit the inversion symmetry, where the former directly comes from the Haldane model,
while the latter acts as a composite symmetry, i.e., $[E||\bar{E}]=[C_2||E][C_2||\bar{E}]$.
It should be emphasized that in the ALM-CI phase, the effective inversion symmetry only exists in the armchair ribbon, because of the presence of the symmetry $[C_2||E]$.
Moreover, this composite inversion symmetry is also reflected by the extremely similar LDOS distribution of its left and right component[cf. Fig.\ref{Armchair-Edgestate}(d)].

\subsection{Transport properties}\label{Transport}

As shown in Figs.\ref{PD-Spinon-Energy} and \ref{Zigzag-Edgestate}, significant spin splitting exists in both the bulk system with periodic boundary conditions
and the ribbon with zigzag periodicity, while it has been shown that the spin splitting is not a sufficient condition for the emergence of
spin current\cite{Fang24}. Therefore, it is natural to ask whether a spin-polarized electric current can be realized in odd-parity ALM materials
which could benefit spintronics applications.

Before discussing transport properties, we first expand the spinon energy dispersion[cf. Eq.\eqref{spinon-disp}],
which is equivalent to the electronic counterpart in the saddle-point approximation,
centered around the Dirac points $\bm{K}$ and $\bm{K}'$ where the low-energy quasiparticle excitations in the ALM-CI phase are accommodated.
In the vicinity of momentum $\bm{K}$ and $\bm{K}'$ with $\bm{k} = \bm{K}^{(')} + \bm{q}$, the momentum-dependent function $\gamma_{\bm{k}}$ and $\gamma_{2\bm{k}}$
can be linearized as $\gamma_{\bm{K}^{(')}+\bm{q}}\approx \tfrac{\sqrt{3}}{2}(iq_x-sq_y)$ and $\gamma_{2\bm{K}^{(')}+\bm{q}}\approx -s\tfrac{3\sqrt{3}}{2}+0(q^2)$
with $s=\pm1$ for valley $\bm{K}$ and $\bm{K}'$, respectively. Therefore the energy dispersion of electron quasiparticles
at around the Dirac points can be approximated as
\begin{eqnarray}\label{approx-disp}
E_{\bm{k}\sigma}^{(\pm)} &\approx& \mu_{\rm eff}+\lambda(Z_{A\sigma}-Z_{B\sigma})(-s\tfrac{3\sqrt{3}}{2}) \nonumber\\
&\pm& \sqrt{ \Delta_{s\sigma}^2 + \frac{3}{4}|tZ\bm{q}|^2 } \nonumber\\
&\approx& \mu_{\rm eff}+\lambda(Z_{A\sigma}-Z_{B\sigma})(-s\tfrac{3\sqrt{3}}{2}) \nonumber\\
&\pm& ( |\Delta_{s\sigma}| + \frac{3}{8|\Delta_{s\sigma}|}|tZ\bm{q}|^2 )=E_{\bm{q}s\sigma}^{(\pm)}\;,
\end{eqnarray}
where $\Delta_{s\sigma}=\Delta_{\sigma}-s\lambda(Z_{A\sigma}+Z_{B\sigma})\tfrac{3\sqrt{3}}{2}$.
From Eq.\eqref{approx-disp}, the characteristics of low-lying electron quasiparticles in the ALM-CI phase can be summarized as:
(i)the presence of the AFM energy gap $\Delta_{\sigma}$ leads to the inversion-symmetry-breaking energy dispersion for electron quasiparticles,
and the direct energy gaps for up- and down-spin electrons are identical, i.e., $2|\Delta_{+1\uparrow}|=2|\Delta_{-1\downarrow}|$;
(ii)the gradient of $E_{\bm{q}s\sigma}^{(\pm)}$ with respect to the momentum $\bm{q}$ vanishes,
indicating that in the presence of energy gap $\Delta_{s\sigma}$,
the linear-in-momentum energy dispersion of low-lying electron quasiparticles in the tight-binding model becomes quadratic-in-momentum,
and thus the low-energy transport properties ought to be dominated by these low-lying electron quasiparticles located around the Dirac points $\bm{K}$ and $\bm{K}'$.

\begin{figure}[h!]
\includegraphics[scale=0.315]{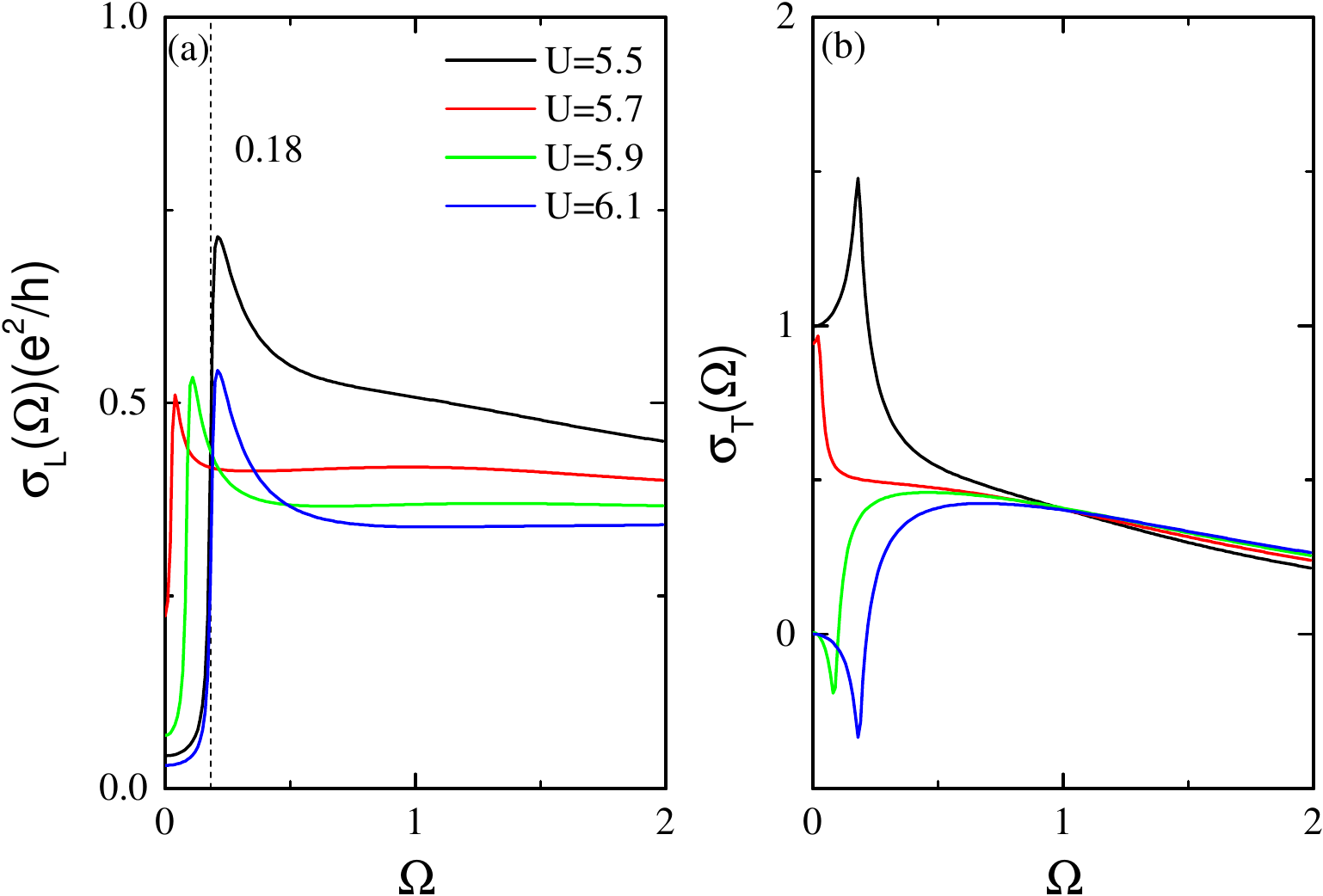}
\caption{(Color online) (a)The longitudinal ($\sigma_{\rm L}$) and (b)the transverse ($\sigma_{\rm T}$) optical conductivity
as a function of energy $\Omega$ at the vanishing hole doping $\delta=0$ and $\lambda=0.3$ with $U=$ 5.5(black), 5.7(red), 5.9(green), and 6.1(blue).
Here, the black dashed line in subfigure(a) denotes the single-particle energy gap $2|\Delta_{+1\uparrow}|$ at $\lambda=0.3$ and $U=$ 5.5.
\label{Optical-Cond}}
\end{figure}
Starting from the Kubo formula[cf. Eq.\eqref{conductivity-1}], we have studied the energy dependence
of the longitudinal($\sigma_{\rm L}$) and transverse($\sigma_{\rm T}$) optical conductivity of HH model in Figs.\ref{Optical-Cond}(a) and (b), respectively,
at the vanishing hole doping $\delta=0$ and $\lambda=0.3$, as well as $U=$ 5.5(black), 5.7(red), 5.9(green), and 6.1(blue).
The main characteristics of optical conductivities, $\sigma_{\rm L}$ and $\sigma_{\rm T}$, can be summarized as:
(i)in the presence of the symmetry $[C_2||\bar{E}]$, there is no spin-polarized electric current even if with significant spin splitting
in the odd-parity ALM phase of HH model, however if the symmetry $[C_2||\bar{E}]$ is broken by external elements such as the sublattice potential,
the spin-polarized electric current will appear\cite{Zeng25-2};
(ii)a pronounced peak appears in the longitudinal and transverse optical conductivity as the energy
$\Omega$ approaches the single-particle energy gap $2|\Delta_{+1\uparrow}|=2|\Delta_{-1\downarrow}|$, which indicates that 
in the presence of the energy gap $2|\Delta_{+1\uparrow}|$, (a)the Drude peak at around $\Omega=0$\cite{Gusynin06,Gusynin07} is transferred to
the energy $\Omega=2|\Delta_{+1\uparrow}|$; (b)the energy dependence of optical conductivities for $\Omega \gtrsim 2|\Delta_{+1\uparrow}|$ can be well
fitted by $\sigma_{L/T}(T,\Omega)=A_{L/T}(1+B_{L/T}/\Omega^2+C_{L/T}/\Omega^4)\theta( \Omega-2|\Delta_{+1\uparrow}| )$ with the fitting parameters
summarized in Table \ref{Fitting-Params},
which is qualitatively consistent with the previous theoretical study\cite{Gusynin07}; In addition, as shown in Table \ref{Fitting-Params},
the larger single-particle energy gap at $U=$ 5.5 and 6.1 significantly increases the coefficient proportional to $\Omega^{-4}$,
which is in contrast to the $2+1$-dimension massive Dirac system with $\sigma(T,\Omega)=\sigma_0(1+B/\Omega^2)\theta( \Omega-2\Delta )$\cite{Gusynin07}.
On the other hand, its high-energy tail differs from this behavior because the higher energy components in the single-particle dispersion set in
that significantly deviates from the linear-in-energy dispersion.
\begin{table}[h!]
\begin{center}
\begin{tabular}{|c|c|c|c|}
\hline
U & $A_L/A_T$ & $B_L/B_T$ & $C_L/C_T$\\
\hline
5.5 & 0.488/0.477 & 0.0317/0.0338 & -0.00046/0.001\\
\hline
5.7 & 0.408/0.486 & 0.00073/0.00095 & -5.12E-7/-1.41E-7\\
\hline
5.9 & 0.353/0.476 & 0.00875/-0.00572 & -2.94E-5/-4.9E-5\\
\hline
6.1 & 0.309/0.455 & 0.0466/-0.0201 & -0.0005/-0.0012\\
\hline
\end{tabular}
\end{center}
\caption{The fitting parameters $A_L/A_T$, $B_L/B_T$, and $C_L/C_T$ of the longitudinal and traverse conductivity, $\sigma_{L}$ and $\sigma_{T}$,
as a function of the onsite Coulomb repulsion $U$ at the Haldane hopping $\lambda=0.3$.
\label{Fitting-Params}}
\end{table}
(iii)the peak position of optical conductivities exhibits a $\lambda$- and $U$-dependence consistent with the single-particle energy gap,
implying that the optical conductivity could be a practical measurement to reveal the topological phase transition experimentally;
(iv)in the topologically nontrivial ALM-CI phase at $\lambda=0.3$ with $U=5.5$ and 5.7, the transverse optical conductivity for each spin exhibits
a quantized plateau, i.e., $\sigma_{\rm T}(\Omega\to 0)=e^2/h$, consistent with the spin-degenerate Chern number $C=1$ in the ALM-CI phase.
However, as the system enters the topologically trivial ALMI phase, the quantized Hall conductivity vanishes, i.e., $\sigma_{\rm T}(\Omega\to 0)=0$.
Therefore, the above results demonstrate that the optical conductivity obtained from linear-response theory
cannot directly reflect the odd-parity spin-splitting form factor, while our recent calculations have demonstrated that
the inversion-symmetry-breaking odd-parity spin splitting in HH model has activated the second-order
spin Hall effect in both the ALM-CI phase and the topologically trivial ALMI phase, which directly comes from the spin-resolved Berry curvature dipole.
The related results on this issue will be presented elsewhere. 
This also reconfirms that the low-frequency linear Hall response in the ALM-CI phase is dominated by the global Chern topology,
rather than by the detailed spin-resolved local redistribution of Berry curvature.

\section{Conclusion}\label{conclude}

In summary, we have shown that the HH model provides a minimal correlated platform for an odd-parity ALM-CI phase.
Using the cluster slave-spin method, we found that the transition from the conventional CI phase into the odd-parity ALM-CI phase
leaves the total Chern number unchanged, but strongly reconstructs the local distribution of the Berry curvature.
The Berry curvature evolves into the spin-valley-locking texture governed
by the symmetry $[C_2||\bar{E}]$; zigzag ribbons develop nonsymmetric chiral edge states; and armchair ribbons remain inversion symmetric.

The linear optical response reinforces this central picture. While the Berry curvature and edge-state spectra are strongly reorganized in the odd-parity ALM-CI phase,
the low-frequency transverse conductivity remains quantized, $\sigma_{\rm T}^{(\uparrow/\downarrow)}(\Omega\to 0)=e^2/h$, and the main spectral features are set by low-energy quasiparticles near the single-particle gap. The odd-parity ALM-CI phase therefore offers a clear example in which the global Hall topology is preserved
even if the local Berry geometry and edge-state structure are substantially reshaped.

Our results place the HH model in the broader context of odd-parity ALM, Hall responses, and quantum-geometric transport\cite{Zeng25_1,Sato24,Fang24,HuM25}.
They also point to experimentally relevant directions, especially in ultracold-atom realizations of the HH model\cite{Zheng14,Jotzu14,Das24}
and in Floquet routes to odd-parity spin splitting in hexagonal antiferromagnets\cite{ZhuTS26,Huang26,Lin25}.
More generally, the present work highlights correlated odd-parity ALMs as promising settings for exploring how symmetry,
interactions, and topology cooperate in Chern systems.

\begin{acknowledgments}

This work was supported by the National Key Re
search and Development Program of China under Grant
 Nos. 2023YFA1406500 and 2021YFA1401803, the Na
tional Natural Science Foundation of China (NSFC)
 under Grant Nos. 12222402, 92365101, 12474151,
 12347101, 12247116, and 12274036, Beijing National
 Laboratory for Condensed Matter Physics under Grant
 No. 2024BNLCMPKF025, the Fundamental Research
 Funds for the Central Universities under Grant No.
 2024IAIS-ZX002, the Chongqing Natural Science Foun
dation under Grants Nos. CSTB2023NSCQ-JQX0024
 and CSTB2022NSCQ-MSX0568, and the Special Fund
ing for Postdoctoral Research Projects in Chongqing
under Grant No. 2024CQBSHTB3156.

\end{acknowledgments}

\vspace{1em}
{\it Data availability}---The data that support the findings of this study are available from the corresponding author upon reasonable request.

\bibliography{BIBHH-Model}

\end{document}